# Tunability of Spin-Dependent Secondary Topological Interface States Induced in an Optical Complex Superlattice

Mengying Hu[⋆], Hui Liu[†] and Shining Zhu

*National Laboratory of Solid State Microstructures, School of Physics,*

*Collaborative Innovation Center of Advanced Microstructures, Nanjing University,*

*Nanjing 210093, China*

**Abstract**

The past decade has witnessed a booming development of topological photonics, which revolutionizes the methodology for controlling the behavior of light. A gigantic achievement is to engineer robust confined modes localized at interfaces between topologically distinct regions, where the optical context can trigger exotic topological phenomena exclusive to photons. Here, we provide an experimentally flexible approach to engineering topologically induced interface states in the visible regime via a unique design of complex superlattice formed by connecting two component superlattices of distinguished topological phases. Assisted by the intrinsic pseudospin degree due to the splitting between TM and TE polarized modes, we attain a precise manipulation of the spin-dependent topological interface states that can manifest themselves straightforwardly through transmission spectra. More specifically, since these topological localized modes stem from the hybridization of artificial photonic orbitals that are of topological origin as well, they are deemed as a novel topological effect and thus named as the secondary topological interface states. Our work develops an innovative and productive strategy to tune topologically protected localized modes, based on which various applications such as selective local enhancement can be exploited.

**Introduction**

Ever since the milestone discovery of the quantum Hall effect in condensed matter lattices, similar topological features have been demonstrated in plentiful systems, including ultracold atoms [1-3], acoustics [4], photonics [5-11], and mechanical structures [12-14], where substantial interest of topological interface states is raised. Serving as robust confined states that exist at interfaces between topologically distinct regions, topological interface states have been intensively studied and opened avenues for utilizations in various fields [15]. Among these, engineering topological interface states in the optical context has captured huge attention in recent years [5,9,16-19], leading to extensive applications in a wide range of platforms and thus serving as a strong driver of current research in topological photonics. Relevant advances of applications encompass local field enhancement [17], topological lasers [20-22] and quantum emitters [19].

The nature of light allows for the exploration of topological phenomena exclusive to photons [23]. When it comes to the manipulation of topological interface states with photons, it is rational to consider harnessing the concept of the pseudospin degree corresponding to the transverse-magnetic (TM) and transverse-electric (TE) polarized modes. In photonics, the pseudospin degree arises from the energy splitting between the mutually orthogonal TM and TE modes[24]. The well-known TM-TE splitting effect can be described in terms of the effective magnetic fields exerting on the photon (pseudo) spin, and hence facilitate us to exploit an applicable approach to realizing the tunability of spin-dependent topological interface states.

On the other hand, recent reports have shown that 1-D superlattice structures can be perceived as a versatile platform to engineer emergent band structure of collective modes [25-30], which are also devoted to pursue the realization of topological interface states [17,31]. In fact, the hybridization of atomic (photonic/acoustic) orbitals accounts for the emergence of these states. However, in these systems, the orbitals themselves are not of topological origin, and thus their existence may be susceptible to perturbations. It is, therefore, urgent for us to explore some robust orbitals that are topologically protected as well. These orbitals originating from band topology hybridize with each other so that the induced topological interface states can be viewed as a sort of secondary topological phenomenon, denoted as secondary topological interface states (STISs), navigating a new pathway to investigate exotic topological phenomena accordingly.

In this work, we firstly bring forward the theoretical concept of STISs and propose an experimentally flexible strategy to engineer spin-dependent STISs in a 1-D optical complex superlattice system, which is deliberately designed by connecting two component superlattices that are topologically distinct. Such STISs, with frequencies lying in the visible regime, reside on the interface between two component superlattices. Specifically, either component superlattice is built up of alternating two photonic crystals (PCs) featured by discriminative band topologies, and hence can host a set of topologically protected interface states. These states supported by the interfaces of PCs act as artificial photonic orbitals, with topological origin confirming the validity of STISs, and are designated as elementary topological interface states (ETISs). Owing to the fact that the appearance of STISs hinges on the polarization of light (pseudospin of photons) in accordance with the TM-TE splitting effect, these STISs are legitimately assorted into two groups on top of the pseudospin degree, each of which can be excited and manipulated independently by means of the incident light with variable in-plane wave vectors. Through meticulous design, the STISs can be liberated from the finite-size effect, pinned at the midpoint of the gaps and observed unambiguously utilizing transmission spectra. Our work develops a creative manipulation mechanism to tune and enhance localized modes, giving an incentive to design novel optical devices equipped with spin control.

*Tunability of ETISs.*——Serving as artificial photonic orbitals, ETISs are situated at the interfaces between two designed (q-type and p-type) PCs. The geometric and dielectric properties of both PCs make them own identical band structure but different topological features, confirming the topological origin of ETISs [32,33] (details in Supplemental Material [34] Section I). Hence, ETISs are robust confined modes protected by topology all along, and exist at all interfaces of the two PCs constructing superlattice structure by stacking alternatively. These ETISs can hybridize with their nearest neighbors, forming collective super-modes and resulting in a superlattice band structure that can be exploited by the tight-binding approach.

Since the on-site resonances and coupling effects of ETISs decide the behaviors of super-modes, the attainability of tunable ETISs is imperative. The fact that ETISs response distinctively to TM and TE polarized excitations should be primarily concerned about. It appends a pseudospin degree of freedom to our photonic system, contributing to strong polarization splittings in both the resonances and coupling frequencies. We show in Fig. 1(a) the splitting effect on resonant frequencies of ETISs obtained by numerical calculations, where we put q-type and p-type PCs (both with 20 unit cells denoted respectively as $q_{20}$ and $p_{20}$) in contact to form an interface supporting ETISs, as illustrated in the upper panel of Fig. 1(a). The ETISs are degenerate in the case of $k_x = 0 \mu m^{-1}$, while such degeneracy is lifted given nonzero $k_x$ cases. To gain insight into couplings within our system, we set up a series of symmetrical structures possessing two interfaces to hold two identical ETISs, based on which the nearest-neighbor coupling effects can be fully investigated. As sketched in the upper panel of Fig. 1(b), each structure is made up of q-type (p-type) PC with certain number (N) of unit cells $q_N$ ($p_N$) sandwiched by two p-type (q-type) PC with 20 unit cells $p_{20}$ ($q_{20}$), and in this context the q-type (p-type) PC plays the role of coupling element separating two interfaces. Therefore, N is perceived as the determinitive factor in terms of the magnitude of coupling strength $J$. We further dig into $J$ through numerical calculations, and reach the conclusion that the sign of $J$ depends on both the parity of N and the type of coupling element. For odd (even) N, $J<0$ ($J>0$) via $q_N$ whereas $J>0$ ($J<0$) via $p_N$, where the former has an identical magnitude with the latter but an opposite sign under the same circumstance. Fig. 1(b) displays the numerical result of $J$ varying with $k_x$ for both polarizations, employing $q_5$ ($p_5$) and $q_6$ ($p_6$) as coupling channels to reveal properties for odd and even N cases (we adopt a subscrip to indicate the number of unit cells of the associated PC). Notice that Fig. 1(b) indicates a conspicuous TM-TE splitting effect on couplings: given any of the four coupling channels, $|J|$ due to TM modes expresses a conversed variation tendency compared with that of TE modes, which rises tardily for the former whereas reduces remarkably for the latter as $k_x$ increases. As a result, the sizable TM-TE splitting effects, which inherently exist in the behaviors of ETISs, are pivotal for us to flexibly engineer super-modes especially STISs appearing in our complex superlattice system.

*Tunability of STISs.* ——We now elaborate on the overall structure of our complex superlattice system shown in Fig. 2(a), which is constructed by stacking two component superlattices on top of each other (call hereafter $\alpha$ and $\beta$ superlattices) and has the ability to support STISs localized at the interface between two components. In our proposed scheme, one of such superlattices ($\alpha$ superlattice) is built up of repeated $p_m$ and $q_n$, whereas the other ($\beta$ superlattice) consists of $p_n$ and $q_m$. $m$ ($n$) designates the number of unit cells of the associated PC, straightforwardly controlling the coupling strength between adjacent ETISs. Thereby, $\alpha$ and $\beta$ superlattices exhibit themselves as analogues to dimerized atomic chains regarding ETISs as photonic orbitals, whose unit dimers are defined respectively as $[p_{m/2}q_n p_{m/2}]$ and $[p_{n/2}q_m p_{n/2}]$ with length of $\Lambda \equiv (m+n)L$, and share identical band dispersions. It follows from Fig. 1(b) that the magnitude of coupling strength correlates negatively with the number of units cells of the PC between adjacent interfaces under the same polarization and $k_x$. We let $m < n$ such that $\alpha$ ($\beta$) superlattice characterized by weak (strong) intra-dimer coupling should be classified into the topological non-trivial (trivial) group. Consequently, the topological distinctiveness of the two phases leads to the formation of exponentially localized interface states sitting the middle of the emergent gap created by super-modes, namely, the STISs. Note that $m$ and $n$ can be chosen as even or odd, so we assign $\lceil m/2 \rceil$ ($\lceil n/2 \rceil$) unit cells to the terminal p-type PCs for $\alpha$ ($\beta$) superlattices to ensure that the interfacial configuration $p_{\lceil m/2 \rceil + \lceil n/2 \rceil}$ only comprises intact unit cells of repetition ($\lceil \ \rceil$ represents the ceiling function introduced to round up to an integer), as depicted in Fig. 2(a).

Then we concentrate our attention on the pseudospin degree of such STISs. Taking account of the TM-TE splitting effects rooted in our system, we rationally extend the original twofold Hamiltonian (regardless of polarizations) to a novel fourfold variant with substantial differences between the two spins (polarizations) owing to the dependence on the in-plane wave vector ($k_x$). Therefore, it is desirable for us to study the polarized properties of STISs. According to the tight-binding theory, the spectral positions of the STISs are exceedingly sensitive to on-site changes while width of the gap and localized length of the STISs are affected by perturbations to the coupling strength. Thus the impressive TM-TE splitting effects we have indubitably corroborated in Figs 1 afford us a viable approach to manipulating spin-dependent STISs host by our complex superlattice system. In experiment, we choose $m=4$ and $n=6$ and our sample is composed of three unit dimers for each component with an open boundary condition (truncated between two unit dimers), fabricated on a glass substrate with the $\beta$ superlattice lying at the bottom. The band dispersions along the stacking direction of these two component superlattices are exhibited in Supplemental Material [34] Section II.

*Tight-binding analysis.* ——Based upon analyses above, we are capable to obtain a comprehensive understanding of our scheme. Accordingly, the full system can be construed by setting up the following tight-binding equations:

$$(\alpha)\begin{cases} (\omega_s - \bar{\omega}_{e\alpha 1,s})\psi_{i,s} = J_{1,s}\psi_{i+1,s} & i=1 \\ (\omega_s - \bar{\omega}_{e,s})\psi_{i,s} = J_{1,s}\psi_{i-1,s} + J_{2,s}\psi_{i+1,s} & i=2,4,\ldots T/2-2 \\ (\omega_s - \bar{\omega}_{e,s})\psi_{i,s} = J_{2,s}\psi_{i-1,s} + J_{1,s}\psi_{i+1,s} & i=3,5,\ldots T/2-1 \end{cases}$$

$$(\text{interface})\begin{cases} (\omega_s - \bar{\omega}_{e,s})\psi_{i,s} = J_{1,s}\psi_{i-1,s} + J_{3,s}\psi_{i+1,s} & i=T/2 \\ (\omega_s - \bar{\omega}_{e,s})\psi_{i,s} = J_{3,s}\psi_{i-1,s} - J_{2,s}\psi_{i+1,s} & i=T/2+1 \end{cases}$$

$$(\beta)\begin{cases} (\omega_s - \bar{\omega}_{e,s})\psi_{i,s} = -J_{2,s}\psi_{i-1,s} - J_{1,s}\psi_{i+1,s} & i=T/2+2, T/2+4,\ldots T-2 \\ (\omega_s - \bar{\omega}_{e,s})\psi_{i,s} = -J_{1,s}\psi_{i-1,s} - J_{2,s}\psi_{i+1,s} & i=T/2+3, T/2+5,\ldots T-1 \\ (\omega_s - \bar{\omega}_{e,s})\psi_{i,s} = -J_{2,s}\psi_{i-1,s} & i=T \end{cases}$$

(1)

where $T$ represents the total number of sites, $i$ enumerates each site with $i=1,2,\ldots,T/2$ ($i=T/2+1,T/2+2,\ldots,T$) assigned for the $\alpha$ ($\beta$) superlattice and the subscripts $s=\uparrow\downarrow$ denote two spins with $s=\uparrow$ ($\downarrow$) for the TM (TE) polarization. The mode amplitude in the $i$th site is given by $\psi_{i,s}$, and $\omega_s$ serves as the reduced frequency. $\bar{\omega}_{e,s} \equiv 1/2(\omega_{e,\uparrow}+\omega_{e,\downarrow})$ represents the average resonant frequency under a specific $k_x$ for ETISs on all sites ($\omega_{e,s=\uparrow\downarrow}$ denotes on-site frequency) except for the first one, which possesses frequency different from $\bar{\omega}_{e,s}$ and is additionally labeled as $\bar{\omega}_{e\alpha 1,s}$. This inequality comes from the fact that the frequencies of ETISs located at the first site are dramatically detuned due to their strong interaction with outside (the air), since we deliberately attach no extra unit cell of p to the leading dimer of $\alpha$ superlattice. In this way, we can obtain purely polarized STISs immune to any other mode, either of which is thus pinned at the ETIS's frequency and can be manipulated independently, otherwise the achievement of pure STISs is elusive due to the inevitable finite-size effect (details in Supplemental Material [34] Section III). Recalling the choise of $m=4$ and $n=6$ in our experiment, we endow the corresponding intra (inter)-dimer coupling strength with $J_{1,s}$ ($J_{2,s}$) for the $\alpha$ superlattice whereas $-J_{2,s}$ ($-J_{1,s}$) for the $\beta$ superlattice stemming from the variation regularity of coupling signs and magnitudes. As for the interface of two components, $J_{3,s}$ is introduced to show the coupling strength between the middle two sites separated by $p_5$.

To get a further step, we harness $J_s$ and $\delta_s$ to describe coupling effects instead of $J_{1,s}$ and $J_{2,s}$ with the definition that $J_{1,s} \equiv -(J_s+\delta_s)$ and $J_{2,s} \equiv J_s-\delta_s$. Therefore, the STIS is supposed to be at the middle of the gap with width of $4|\delta_s|$. Incorporating the detuning of the leading site that owns a magnitude on the order of 8THz into the scheme, we achieve the reduced frequency spectra evolving with $\delta_s$. Fig. 2(b) illustrates the case of both TM and TE super-modes for a complex superlattice of 40 dimers (20 for each component) provided $k_x = 0\mu m^{-1}$. The two spectra are virtually overlapped due to degeneracy, either of which reveals 80 super-modes within the spectral range of interest. Shown definitely in Fig. 2(b), STISs settle at the center of the gap featured by $\omega_s = \bar{\omega}_{e,s} = \omega_{e,\uparrow} = \omega_{e,\downarrow}$ regardless of polarizations. When $k_x \neq 0\mu m^{-1}$, the degeneracy of TM and TE super-modes is removed. Fig. 2(c) (2(d)) depicts the spectra of TM (TE) super-modes when $k_x = 5\mu m^{-1}$ for the same configuration as in Fig. 2(b), from which we can see the STIS possesses the frequency of $1/2(\omega_{e,\uparrow} - \omega_{e,\downarrow})$ ($-1/2(\omega_{e,\uparrow} - \omega_{e,\downarrow})$). The modes next to the boundary of the upper band is attributed to the detuning term (details in Supplemental Material [34] Section III). $\delta_s$ corresponding to the experimental structure are marked by dashed lines in Fig. 2(b-d), along with electric profiles of associated STISs (insets), convincingly testifying our former argument.

*Experimental demonstration.* —— Our sample consists of 6 dimers (3 for either *α* or *β* superlattice) in order to ensure the implement of experiment, within which the super-modes still preserve typical properties demonstrated for the 40 dimers system. Details of the materials and the thickness of the cells comprising the complex system are offered in Supplemental Material [34] Section I. These states can be excited easily by the incidence of plane-wave light, and detected immediately utilizing transmission measurements. Fig. 3(a) presents theoretical values of reduced frequencies as $k_x = 0\mu m^{-1}$, where states with identical frequencies appear in pair due to degeneracy of polarizations. The two zero-mode STISs can be observed in the transmission spectrum under normal incidence, shown in Fig. 3(b) as a sharp peak in the middle of the gap. As the incident light becomes oblique, the splitting effects should be taken into account. Figs. 3(c) and 3(d) exhibit reduced frequencies and transmission spectra corresponding to an incident angle of $36°$ for both polarizations. The TM-TE splitting effects give rise to the removal of degeneracy so that the STIS of TM (TE) polarization blueshifts (redshifts), confirmed by the conformance between theoretical (3(c)) and experimental (3(d)) results. Notice that the transmission spectra imply the excitations of pure STISs, which remain fixed at the midgap frequencies and hence are impervious to the finite-size effect through our ingenious design. In addition to the positions of STISs, the widths of gaps and the distributions of extended modes seen in Fig. 3(b) and 3(d) are in consistency with those calculated in Fig. 3(a) and 3(c). All of these are sufficient to evidence the efficacy of our tactics to tune spin-dependent STISs.

The intrinsic splitting in frequency of TM and TE super-modes originates from the wavevector-dependent effective field exerting on the polarization of photons. A similar behavior arises for electrons in a Zeeman magnetic field. Thereby, the anisotropy between TM and TE polarized STISs is gradually pronounced with the enhancement of the "magnetic field" (that is, the increase of $k_x$). To verify this, we continuously measure the transmission spectra as the incident angle varies from $0°$ to $60°$ for both polarizations utilizing the spectrometer with a broad detection range and high precision (no more than 0.5 degree), and the results are exhibited in Fig. 4. Besides, in Fig. 4 we also plot the locations of STISs acquired by tight-binding methods (white dashed lines) and numerical calculations (red solid circles), matching great well with experiments.

*Conclusion and discussion.* —— We research on the exotic topological states of STISs theoretically and experimentally within a complex superlattice system. We deliberately create topologically protected photonic orbitals (ETISs) and realize controllable hybridizations, which acts as indispensable elements to achieve STISs. Drawing on the pseudospin degree of photons, we demonstrate the high-precision tunability of spin-dependent STISs, arising from the reliance of both the on-site resonance of single ETIS and coupling effects between nearest-neighbor ETISs on the pseudospin degree. In our study, the robust STISs work at visible frequencies, whose facile and precise manipulation holds great promise in the development of visible light applications, such as visible light communication devices [35] and local fluorescence enhancement [36]. Moreover, it is noteworthy that each STIS displays itself as a purely polarized interface mode immune to the finite-size effect, characterized by the excited frequency identical to that of the ETIS. Therefore, STISs with various frequencies under identical excited condition are readily available by tuning the on-site resonance of ETISs continuously [30], where the existence of ETISs is always protected by topology and hence endows us with more freedom to manipulate STISs. Our complex superlattice system serves as a new platform to investigate intriguing topological phenomena in which the TM-TE splitting effect counts, inspiring us to introduce the mechanism of spin-orbit coupling [37,38] in the ensuing work so as to enrich inherent physics and potential applications. Besides, the obtainability and adjustability of STISs render us a generic strategy expected to be extended in 2-D or 3-D optical structure, and may facilitate applications inclusive of spin-control laser emitters and optical devices [39-42]. Furthermore, what makes our system more appealing is that it can be exploited for the exploration on some frontiers of topological photonics. For instance, by rationally harnessing materials with absorption or with intensity-dependent refractive indexes to tune the STISs' eigenfrequencies, non-Hermitian [43] or nonlinear phenomena associated with secondary effects are expected to be investigated.

*Acknowledgements.*——  M. Hu. thanks K. Ding for helpful discussions. We gratefully acknowledge the support of the National Key Projects for Basic Researches of China (Grants No. 2017YFA0205700 and No. 2017YFA0303700), and the National Natural Science Foundation of China (Grants No. 11690033, No. 61425018, No. 11621091, and No. 11374151).

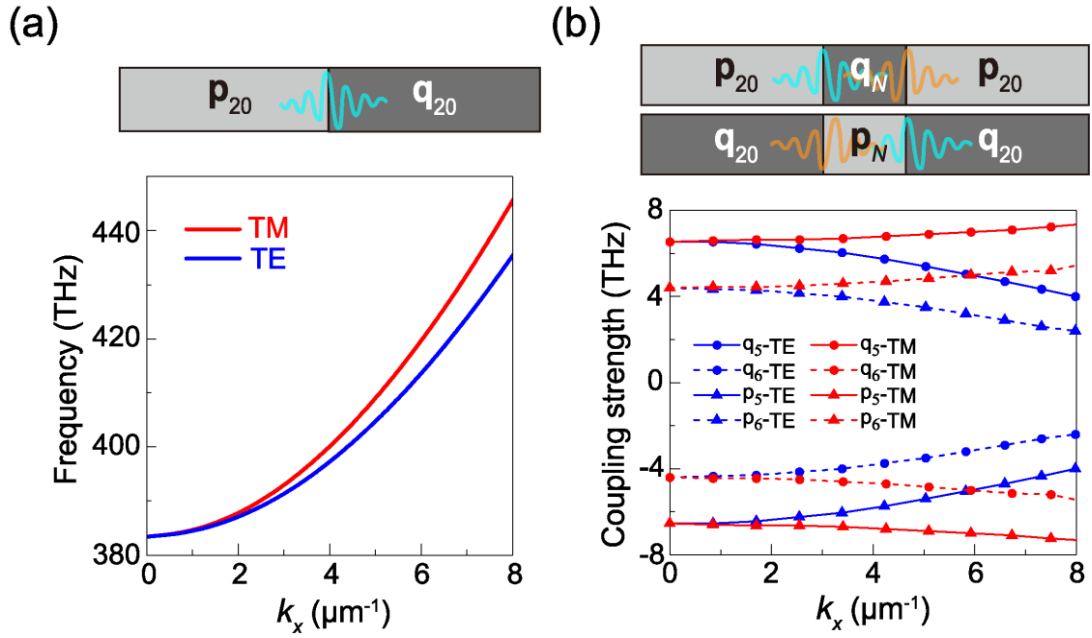

**FIG. 1** (a) The in-plane dispersion relations for TM and TE polarized ETISs. (b) Dependence of the coupling strength on the in-plane wave vector $k_x$ calculated for both polarizations. The number of unit cells of the q (p)-type PC designated as the coupling channel for either TM or TE polarized excitation is labeled by the subscript. Sketches of designs to obtain single ETIS, coupling modes between adjacent ETISs are depicted in the upper panels of (a) and (b), respectively.

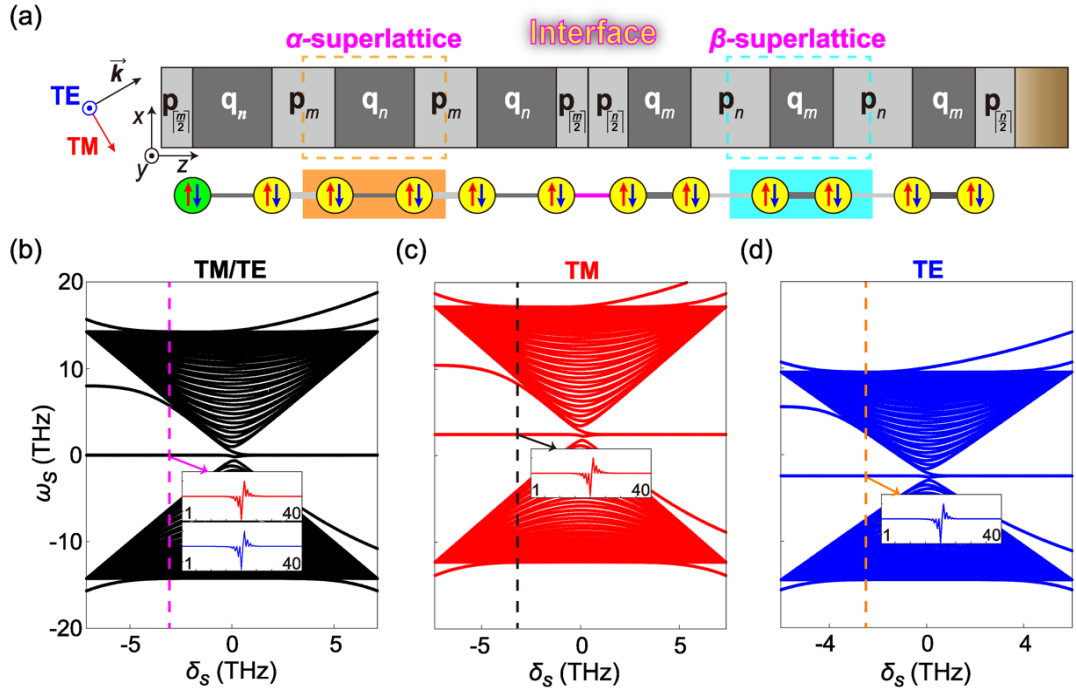

**FIG. 2** (a) Structure diagram of the complex superlattice (upper panel) and associated schematic illustration (lower panel). In the lower panel, each from the second to the terminal sites is indicated by a yellow circle while the first site is portrayed by a green one, with a red upward (blue downward) arrow inside showing the spin of TM (TE) polarization. The coupling strength pertaining to q (p)-type PC is represented by a dark (light) grey line. Thick and thin lines refer to strong and weak couplings, respectively. (b-d) Reduced frequency spectra as a function of $\delta_s$ under TM and TE polarizations with (b) $k_x = 0\,\mu m^{-1}$ and (c-d) $k_x = 5\,\mu m^{-1}$ for a finite chain composed of $\alpha$ and $\beta$ superlattices stacking on top of each other, either of which comprises 20 dimers. Arrows indicate the positions of STISs as $m = 4$ and $n = 6$, where the insets alongside reveal their associated distributions of displacements.

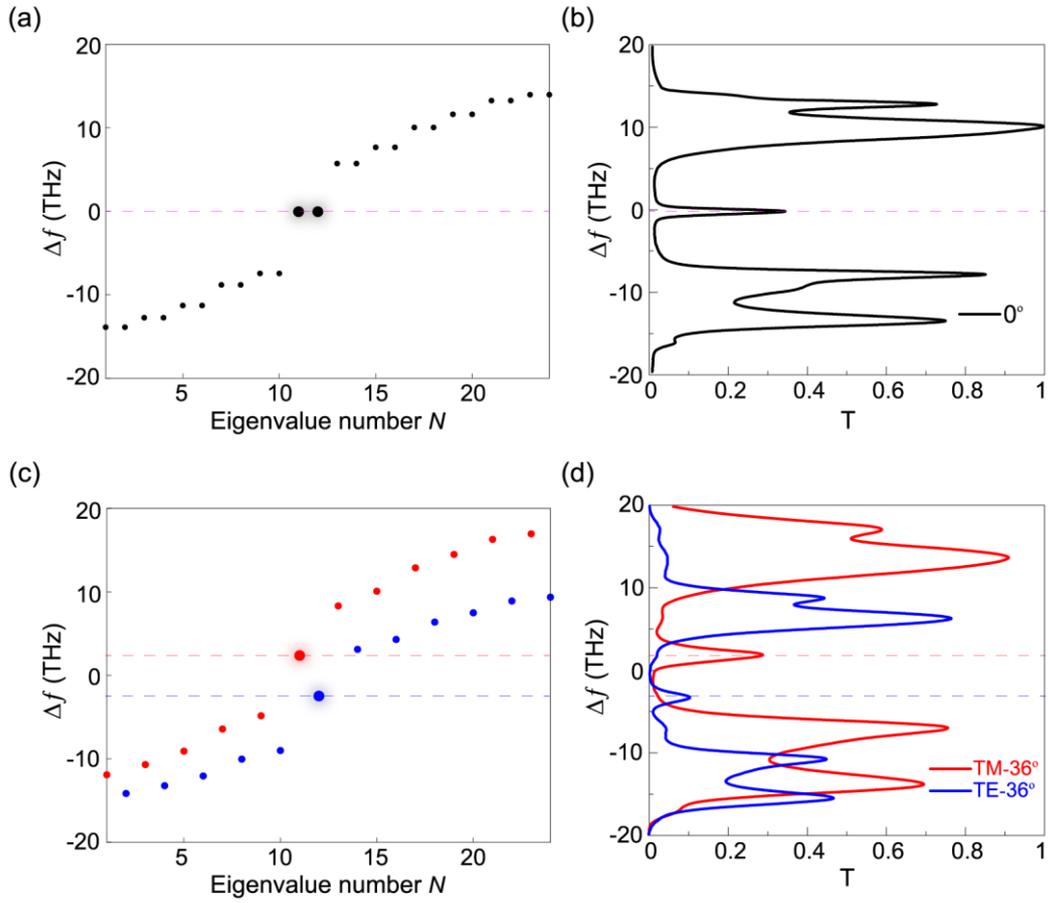

**FIG. 3** Reduced frequencies deduced from the tight-binding approach and measured transmission spectra for the sample without TM-TE splitting (normal incidence) (a, b), and with TM-TE splitting (oblique incident angle of 36°) (c, d). For clarity, markers of STISs are deliberately magnified in (a) and (c), corresponding to the peaks for STISs sitting in the middle of gaps that are emphasized by dashed lines in (b) and (d), respectively.

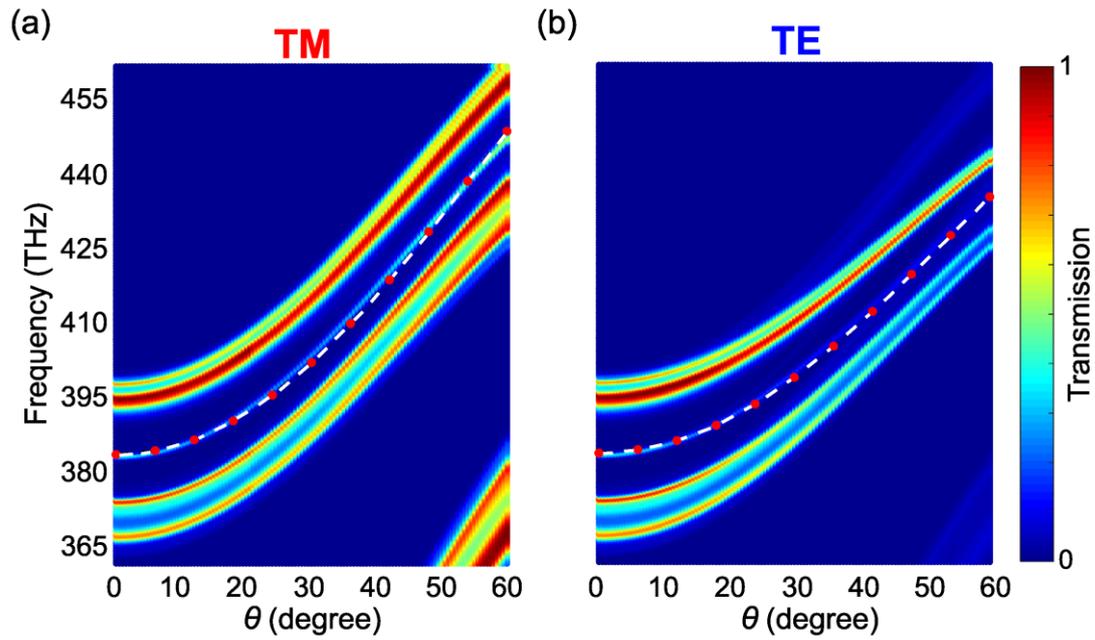

**FIG. 4** Measured transmission spectra as the incident angle alters from $0°$ to $60°$ under (a) TM and (b) TE excitations. White dashed lines and red solid circles display the trajectories of STISs achieved by tight-binding methods and numerical calculations utilizing COMSOL, respectively.